\documentclass[acmsmall]{acmart} 

\usepackage{graphicx}
\usepackage{multirow}
\usepackage{amsmath,amssymb,amsfonts}
\usepackage{textcomp}
\usepackage{booktabs}
\usepackage{algorithm}
\usepackage{algorithmicx}
\usepackage{algpseudocode}
\usepackage{listings}
\usepackage{url}
\usepackage{hyperref}
\usepackage{todonotes}

\definecolor{codegreen}{rgb}{0,0.6,0}
\definecolor{codegray}{rgb}{0.5,0.5,0.5}
\definecolor{codepurple}{rgb}{0.58,0,0.82}
\definecolor{backcolour}{rgb}{0.95,0.95,0.92}

\lstdefinestyle{mystyle}{
    backgroundcolor=\color{backcolour},   
    commentstyle=\color{codegreen},
    keywordstyle=\color{magenta},
    numberstyle=\tiny\color{codegray},
    stringstyle=\color{codepurple},
    basicstyle=\ttfamily\footnotesize,
    breakatwhitespace=false,         
    breaklines=true,                 
    captionpos=b,                    
    keepspaces=true,                 
    numbers=left,                    
    numbersep=0pt,                  
    showspaces=false,                
    showstringspaces=false,
    showtabs=false,                  
    tabsize=2
}
\renewcommand\footnotetextcopyrightpermission[1]{} 


\begin{document}

\title{Context-Aware Operational Security for Autonomous Drones}

\author{Burak Tufekci}
\email{burak.tufekci@unt.edu}
\affiliation{%
  \institution{University of North Texas}
  \department{Department of Computer Science and Engineering}
  \city{Denton}
  \state{Texas}
  \country{USA}
}

\author{Cihan Tunc}
\email{cihan.tunc@unt.edu}
\affiliation{%
  \institution{University of North Texas}
  \department{Department of Computer Science and Engineering}
  \city{Denton}
  \state{Texas}
  \country{USA}
}

\begin{abstract}
Autonomous drone-based services have been gaining significant interest across various application domains due to their mobility, flexibility, cost-effectiveness, and capability to integrate various sensors and actuators. However, operational failures or cyberattacks targeting drone systems can lead to severe economic impacts and safety concerns. Hence, ensuring secure and reliable autonomous drone operations is essential for the safe deployment of drone-enabled services. 
Nevertheless, the traditional security measures fall short due to drones' limited computational resources and power budget (battery), as well as the temporal and sequential behavior of drone operations due to drones' mobile nature. In this paper, we address this gap by utilizing Recurrent Neural Networks (RNNs), specifically focusing on Long Short-Term Memory (LSTM) networks, for autonomous drone operation security and reliability due to their temporal and sequential analysis capabilities. We leverage these capabilities for anomaly detection in autonomous drone sensor data and operation commands, which we refer to as Denial of Usage Detection Engine IDS (DUDE-IDS). We integrated the proposed DUDE-IDS into the drone mission computer (i.e., operating directly on drones rather than an edge node or ground control stations) to monitor data flows and to detect potential threats in real-time. Extensive experimental results demonstrate the effectiveness of this approach in identifying anomalies associated with GPS spoofing, Man-in-the-Middle (MITM), replay, and Denial-of-Service (DoS) attacks with 98\% accuracy. We also evaluate our resource utilization and power consumption under different configurations, demonstrating the applicability of our approach in active drone operations in real-time. 
\end{abstract}


\maketitle

\section{Introduction} \label{Introduction}

In the rapidly evolving landscape of autonomous systems, drones have emerged as critical components across various domains, e.g., surveillance~\cite{dapp:Huang}, agriculture~\cite{dapp:Rejeb}, delivery~\cite{dapp:Das}, and environmental monitoring~\cite{dapp:Kyrkou}. The global market for autonomous drones was valued at USD 15.9 billion in 2023 and is expected to grow to USD 53.4 billion by 2030, with an estimated Compound Annual Growth Rate (CAGR) of 18.9\% from 2024 to 2030~\cite{stats:autodrone}. 
As autonomous drones perform tasks without human intervention, they require onboard sensors and actuators along with algorithms to navigate, make decisions, and adapt to changing conditions. However, drone operation security is a major challenge due to the large set of potential threats and vulnerabilities~\cite{drone:vulnerability}. 
Without robust security measures, drones are susceptible to attacks such as unauthorized access, data breaches, manipulation of sensor data, or even a complete control takeover. 
For example, autonomous drones rely heavily on GNSS (Global Navigation Satellite Systems) for navigation, positioning, and mission execution, but an intercepted signal can cause failure in operations~\cite{rigoni2024towards,michieletto2022robust}. 
Additional cases have been reported in recent years: A contractor inserted a cyber-time bomb in drone goggle firmware causing the failure at a predefined time~\cite{icsstriveContractorInserts}, a drone supply chain has been breached affecting human safety~\cite{RussianGoggles26}, and so on in addition to the communication/network-based cyberattacks~\cite{drone:vulnerability,tufekci2024dude}. 
Therefore, ensuring comprehensive drone operational security is essential to protect and maintain safe and reliable autonomous drone missions (for the rest of the text, drone will be used to refer to autonomous drones mostly, to reduce the repetition).

Conventional security mechanisms (e.g., anti-virus software, access control) are often insufficient to protect due to the nature of drones (limited resources, computational capabilities, and battery, along with their mobility and connectivity). 
Alternatively, Machine Learning (ML) can offer advanced capabilities to detect anomalies, predict potential threats, and adapt to evolving security landscapes~\cite{ids:Chaabouni, ids:Zolanvari}. 
A critical aspect of drone operation is the analysis of \textit{sequential sensor data}, especially for the drone control mechanism (i.e., flight controller)~\cite{konapalli2025reverse}. Consider an autonomous surveillance drone tasked with monitoring a restricted area. This drone relies on continuous input from its onboard sensors (such as GPS, cameras, and Inertial Measurement Units (IMUs)) to navigate and perform its mission effectively. Detecting anomalies (due to unexpected movements, failures in any of the devices, tampering attempts, or cyberattacks targeting the drone's control, for example) requires analysis of its sequential sensor data. If the drone encounters any GPS signal spoofing, data patterns from multiple sensors will show inconsistencies, which can be used to alert and consider alternative/fail-safe solutions (e.g., other navigation methods, returning to a safe location); otherwise the successful attack can cause the failure of the operation, potentially causing drone to fly to another destination or even crash.

It has been observed that commonly used ML approaches are not well-suited for analyzing drones' sequential data as they treat each data point independently and lack the ability to maintain context over time~\cite{ids:Hwa}. For example, Feedforward Neural Networks (FNN)~\cite{feedforward:Hornik} process each input independently, missing critical temporal patterns. While decision trees~\cite{decisiontree:song} are effective for static data, they struggle with dynamic sequences. Even though Convolutional Neural Networks (CNNs)~\cite{ccn:Li} are excellent for spatial data, they require significant adaptation to capture temporal dependencies. Though transformers~\cite{transformers:Khan} are powerful for sequence modeling and capable of capturing long-range dependencies, they demand large datasets and extensive computational resources, making them less practical for real-time drone operations, considering limited drone resources -- this is especially the concern for on-drone solutions because edge computing or Ground Control Station (GCS)-based solutions cannot provide real-time and reliable solutions. To address this problem, Recurrent Neural Networks (RNNs)~\cite{rnn:medsker} can be a great alternative as they are particularly well-suited for such tasks, as they are designed to recognize patterns in sequences by maintaining a form of memory over time and capturing temporal dependencies and contextual information. Hence, RNNs can provide a powerful framework for anomaly detection systems for potential drone failures or security threats. Based on this motivation, the contributions of this paper are as follows:

\begin{itemize}
	\item We define drone operational states using a novel data structure based on a state transition table to define individual operating points.
	\item We created a labeled dataset using a DIY drone under cyberattacks targeting MAVLink (Micro Air Vehicle Link~\cite{mav:Nguyen}) protocol-based command and sensor readings. The existing literature lacks such a public dataset, and this dataset would greatly benefit the research community. Therefore, we will share this dataset on our research website~\cite{smartcyberspaceSmartCyber}. 
	\item We propose the LSTM-based (extended) Denial of Usage Detection Engine IDS (DUDE-IDS) focusing on \textit{sensor data and received commands} to ensure drone operation safety.
	\item We implement and demonstrate the proposed DUDE-IDS on a physical drone for real-time detection, demonstrating its superiority and suitability for this problem. 
    \item We also compare our approach with a CNN implementation using different hardware environments to demonstrate the effectiveness and applicability of the proposed DUDE-IDS. 
\end{itemize}

The remainder of this paper is organized as follows: 
Section~\ref{sec:RelatedWork} reviews the literature focusing on drone communication threats. Section~\ref{sec:Methodology} describes the operational states of drones and explains DUDE-IDS architecture with its features. Section~\ref{sec:ExperimentalSetup} explains our experimental setup and Section~\ref{sec:ResultsandDiscussions} discusses experimental results. Finally, Section~\ref{sec:Conclusion} summarizes and concludes the paper.

\section{Related Work} \label{sec:RelatedWork}

\subsection{Sequential Data in Drones}

Drones are equipped with various sensors for continuous monitoring of their environment and operations such as Global Positioning System (\textit{GPS}) and Global Navigation Satellite System (\textit{GNSS}) for geographic coordinates and altitude, \textit{accelerometers} for velocity and acceleration in three dimensions, \textit{gyroscopes} for angular velocity and rotational movements, \textit{barometers} for atmospheric pressure changes, and \textit{magnetometers} for magnetic fields, which all assist in navigation by providing critical heading data. 

MAVLink is a lightweight application-layer communication protocol commonly used for communication between drones and GCSs, and drone components~\cite{mav:Nguyen}. 
MAVLink commands can be grouped as follows: 
    (a) \textbf{Flight Mode Changes: } Sequences of commands to switch between different flight modes (e.g., manual, auto, or return-to-launch). 
	(b) \textbf{Waypoint Instructions:}  A series of commands that guide the drone from one point to another as a flight path. 
	(c) \textbf{Parameter Adjustments:} Commands to change the flight parameters such as speed, altitude, or heading. 
	(d) \textbf{Mission Commands:} Instructions for complex operations, such as survey missions or payload drops. 
For more information about MAVLink and related threats, please refer to~\cite{koubaa2019micro, tufekci2024enhancing}. 

A sequence of waypoint commands transmitted using MAVLink defines a specific flight path, and any anomaly or deviation from this sequence can lead to mission failure or potential hazards. 
Similarly, sensor data from drones provide sequential data with temporal and/or spatial dependencies that can be affected by various effects (including but not limited to cyberattacks, environmental conditions, and failures/faults in drone components and networks).  
Consider a surveillance drone navigating a predefined path, which records the altitude at time $t_{1}$ $100$ meters, at time $t_2$ as $105$ meters with an additional detection of a strong upward gust of wind. In such a case, the next altitude would be highly influenced by the previous and current altitude record and wind. 
Assuming the wind persists (with no change in the other control effects), the altitude at the future time point  $t_3$ will likely be around $110$ meters. However, a major difference of altitude at $t_3$, such as $50$ meters, represents an unexpected behavior that might be due to an attack, improper sensor/command/actuator operation, or a failure. 
Therefore, for successful drone operations, sequential data emerging from sensor readings and MAVLink commands are two main independent sources that should be investigated continuously.

\subsection{Cyber-Threats Targeting Drone Operations}\label{sec:threats}

Drone security has been emphasized starting with threat analysis. 
Krichen et al. studied the security challenges of drone communications, including possible threats, attacks, and countermeasures~\cite{dsecurity:Krichen}. The authors categorized attacks against drones into four categories: Drone-To-Ground Station, Drone-To-Network, Drone-To-Satellite, and Drone-To-Drone, and identified the main vulnerabilities, including insecure communication protocols, potential software and hardware malfunctions, and the risk of hacking and malicious use. They suggested using Blockchain technology, ML techniques, fog computing, and Software Defined Networks (SDN). Similarly, Nguyen et al. gave an overview of the current drone threats with possible countermeasures in three main categories~\cite{dsecurity:Nguyen}: vulnerabilities of devices integrated into drones, communication links, and privacy issues. They emphasized sensor security importance for information gathering and flight safety and suggested using multiple sensors instead of a single sensor. The authors also highlighted Wi-Fi vulnerabilities and suggested countermeasures (e.g., game-theoretic countermeasures, using hash functions, and hardware sandboxing). Additional studies focusing on threat modeling for drones by Salamh et al.~\cite{dsecurity:Salamh} and Iqbal et al.~\cite{dsecurity:Iqbal} aimed for confidentiality, authenticated access, software/data integrity, system availability, and accountability. 

Another major concern for the security of drones has been the Global Positioning System (GPS) and Global Navigation Satellite System (GNSS) spoofing attacks targeting drone position and velocity causing incorrect paths. 
Davidovich et al. aimed at detecting GPS spoofing by analyzing camera video stream~\cite{dsecurityspoofing:Davidovich}. The authors evaluated their approach using a dataset of drone camera videos and demonstrated their approach with high accuracy. Michieletto et al.~\cite{dsecurityspoofing:Michieletto} proposed secure navigation under GNSS spoofing by leveraging both GNSS and inertial sensors using (1) GNSS spoofing detection through ML using the received GNSS signal characteristics, (2) inertial-based localization using an Extended Kalman Filter (EKF) to estimate position and velocity, and (3) GNSS-aided localization. 
The proposed method was evaluated using real-world GNSS data and showed improved robustness against GNSS spoofing. Another study, by Sung et al.~\cite{dsecurityspoofing:Sung}, aimed at detecting GPS anomalies using GPS signal strength and satellite positioning information using a 1D-CNN. 

Additional security concerns have been studied, including a lightweight digital signature for a secure flight controller communication protocol to prevent MITM attacks~\cite{dsecuritymitm:Li}. 
Piggott et al. utilized Net-GPT, a malicious chatbot engineered to understand network protocols and execute MITM attacks on drones~\cite{dsecuritymitm:Piggott}. Using an edge server utilizing fine-tuned large language models (LLMs), they mimicked network packets from a public network traffic repository with the potential of augmenting traffic between a drone and a GCS. The high generative capacities of LLMs allow for crafting network packets that align with the existing communication context. Their experimental results indicate that LLMs are effective tools for adversaries, demonstrating that models such as Llama-2-13B and Llama-2-7B achieve prediction accuracies of 95.3\% and 94.1\%, respectively. To avoid replay attacks, Omar et al. proposed intercepting and retransmitting control signals using Software-Defined Radios (SDRs)~\cite{dsecurityreplay:Omar}. The system utilizes RF signal scanning to detect the presence of a drone within a certain geographical area and then records the control signals. They also discuss using encryption and authentication to prevent replay attacks. Sanchez et al. investigated a sine wave with time-varying frequency as an authentication signal for detecting replay attacks on drones~\cite{dsecurityreplay:Helem}. 
Abdulwahhab et al. focused on flooding attacks against Flying Ad-hoc Networks (FANETs) as mobility networks for drones~\cite{dsecuritydos:Abdulwahhab}. The authors proposed a threshold-based detection and mitigation mechanism based on the Ad Hoc On-Demand Distance Vector (AODV) routing protocol. 

Furthermore, Hassler et al. proposed a cyber-physical IDS for drones to address growing cyber threats~\cite{anomaly:Hassler}. They claimed that current IDSs are limited by their reliance on either cyber or physical features alone, which fail to capture the full scope of drone vulnerabilities. They introduced a novel approach that fuses both cyber and physical data for improved detection accuracy. Their testbed was developed to simulate various cyberattacks (e.g., DoS, replay, evil twin, false data injection) on drones, and they publicly shared the corresponding datasets. Their study evaluated ML models such as support vector machines, neural networks, and convolutional networks, analyzing how feature fusion enhances detection. The experimental results demonstrated that fused data models outperform single-domain models, offering generalization across attack complexities and improved protection for drone operations.

\textit{To the best of our knowledge}, \textit{the proposed DUDE-IDS solution for sensor and control} is one of the pioneers in the operational safety of drones. While several studies have explored various aspects of drone safety and anomaly detection (e.g.,~\cite{safety:Aldos, safety:Lim, safety:Rakotonarivo, safety:Maria, anomaly:Akram, anomaly:Alzahrani, anomaly:Tran}), the specific focus on utilizing MAVLink commands and sensor readings to enhance the operational safety of drones remains largely unexplored. Furthermore, several studies did not evaluate the applicability of the proposed solutions on physical drones with real-time (or near real-time) capabilities. Moreover, publicly accessible control datasets for drones are limited, and the existing studies with available data primarily focus on network security rather than operational security. \textit{These gaps highlight the novelty and significance of our approach} in drone command and sensor-based options using real-time monitoring and anomaly detection.
We specifically focus on detecting the following drone cyber-threats in our approach. 
(a) \textbf{Sensor Spoofing:} This threat involves manipulating drone sensors to provide false data, potentially leading to incorrect state transitions or mission failures. 
(b)	\textbf{Man-in-the-Middle (MITM):} A threat where an attacker intercepts or alters the communication between a drone and its controller compromising command integrity and state transitions. 
(c)	\textbf{Replay Attacks:} Legitimate commands or data packets are maliciously repeated, potentially causing harmful actions or transitioning into unintended states. 
(d)	\textbf{Denial of Service (DoS):} Normal drone operations are disrupted by overwhelming with fake commands or data, leading to state transitions that can interrupt or compromise its mission. 
We selected these attacks because they are commonly observed attacks both in drone and IoT-related studies. 
\section{Methodology}\label{sec:Methodology}

\begin{figure}[htbp]
 \centering
	\includegraphics[width=.45\columnwidth]{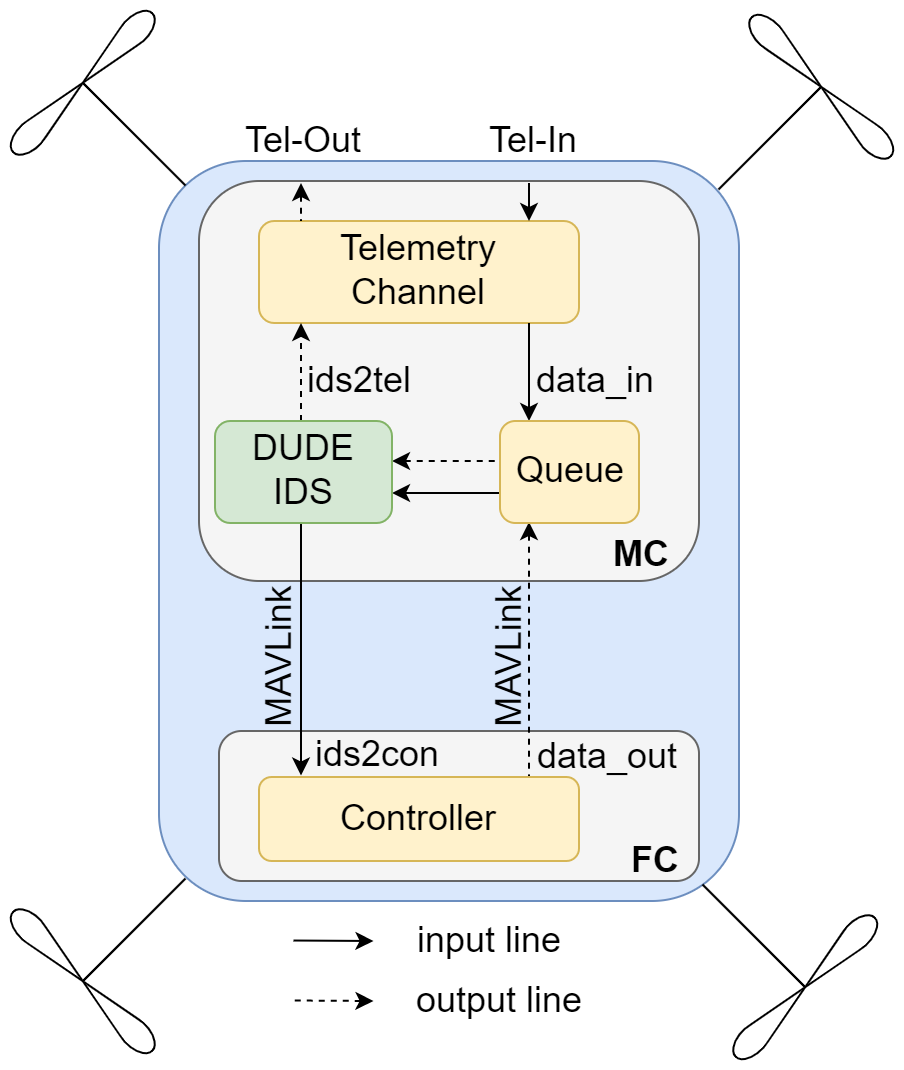}
	\caption{Architecture of the proposed DUDE-IDS solution on a physical drone.}
	\label{fig:architecture}
\end{figure}

The high-level architecture of the proposed \textit{DUDE-IDS for sensor and control protection }is illustrated in Figure~\ref{fig:architecture}. A drone is typically equipped with a \textit{Flight Controller (FC)} that is responsible for the control and stabilization of flight components in real-time with a direct connection to motors, sensors (e.g., GPS, IMU, barometer), and other actuators (e.g., electronic speed controllers and servo motors). The \textit{controller} module on the FC is responsible for executing flight control algorithms, managing the drone's flight dynamics, and processing commands. Due to limited resources of the FC, complex operations (e.g., analysis/control of additional sensors and actuators and advanced tasks such as object detection/recognition) are generally processed by a more sophisticated control unit \textit{Mission Computer (MC)} (e.g., using Raspberry Pi or NVIDIA Jetson Nano) and send to the FC control unit. 
MC also hosts a telemetry channel module to manage incoming (Tel-In) and outgoing (Tel-Out) telemetry data. The MC and FC communicate with each other using MAVLink to exchange mission data (\textit{ids2con}) and flight status information (\textit{data\_out}).

The proposed \textit{DUDE-IDS is strategically placed on the MC} instead of on the FC (due to limited FC capabilities) to monitor and analyze the command traffic between the MC and FC and to detect potential threats or anomalies. Upon detecting a threat/anomaly, DUDE-IDS creates an alert, drops the requested command or data, and sends a command to the FC to take immediate action, such as changing the flight path or initiating safety protocols (e.g., return-to-launch, emergency landing, communication lockdown). 
For the successful operation, we also introduced a \textit{Queue} as intermediary storage and initial processing of telemetry and controller data to prevent performance bottlenecks. 

Imagine a drone delivery company that uses drones to transport packages in a metropolitan area. During a routine delivery mission, a malicious actor attempts to hijack the drone by injecting unauthorized commands using an MITM attack, which instructs the drone to deviate from its intended flight path and fly to the attacker's destination. DUDE-IDS detects this anomaly by identifying the deviation from the predefined safe flight path and operational command patterns. Then, it generates an alert, drops the unauthorized command to prevent the FC process, and sends a safe command to the FC to initiate immediate safety actions. These actions may include switching the drone to a return-to-launch mode and altering the flight path to avoid potential threats. This response needs to be quick to ensure the safety of the drone, its operation, and its payload.

\subsection{Operational States of Drones}\label{sec:states}
\begin{figure}[ht]
	\centering
	 \includegraphics[width=0.7\columnwidth]{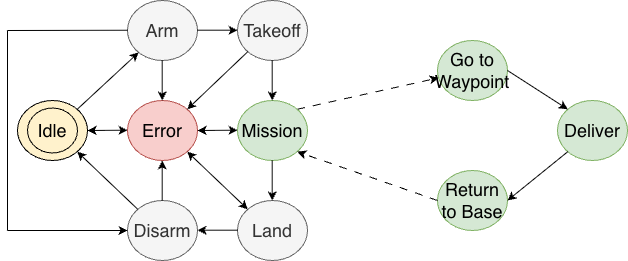} 	
	\caption{State machine for fundamental drone operations. The mission can have further individual sub-states (shown with dotted-lines).}
	\label{fig:StateMachine}
    \Description{State machine for fundamental drone operations. The mission can have further individual sub-states.}
\vspace{-3mm}
\end{figure}

For operational state analysis, first, we created a state machine representing fundamental drone operations in a structured and systematic way, as shown in Figure~\ref{fig:StateMachine}. Consider a drone deployed for package delivery. In the initial state (\textit{Idle}), the drone is only powered on without any activities. Then, the next operational states for such a task would be \textit{Arm} (enabling flight operations), \textit{Takeoff}, and \textit{Mission}, where \textit{Mission} can be further detailed with \textit{Go to Waypoint}, \textit{Deliver the Package}, and \textit{Return to Base} states depending on the operational requirements (the detailed states have been briefly demonstrated). Finally, \textit{Land} and \textit{Disarm} states are applied. If during the operation of any of these states, an incorrect/abnormal command or an attack occurs, the drone can move to the \textit{Error} state, following with options of \textit{Mission}, \textit{Land}, and \textit{Idle} states. Suppose the drone encounters a GPS spoofing attack during the \textit{Go to Waypoint} state (\textit{Mission}), causing it to deviate from its intended path, it can transition to the \textit{Error} state and revisit the \textit{Mission} state for \textit{Return to Base}.

We detail the state transitions using Table~\ref{tab:StateTransition} to provide further information including \textit{Current State} with the \textit{Event/Condition} in which each transition occurs (i.e., \textit{Next State}) and the associated \textit{security threats} (aiming sensor spoofing (S), man-in-the-middle attacks (M), replay attacks R, and DoS/DDoS (D)). 
By leveraging the state transition table, the proposed DUDE-IDS ensures that the drone operates within predefined safety parameters, thereby mitigating potential accidents and ensuring successful mission completion. This structured approach is crucial for managing the complexity of various operational scenarios.

\begin{table*}[hbtp]
    \caption{State transition table.}
    \centering
    \label{tab:StateTransition}
    \small 
    \begin{tabular}{p{2cm} p{3.5cm} p{1.5cm} p{3.5cm} p{1.5cm}}
        \hline
        \textbf{Current State} & \textbf{Event/Condition} & \textbf{Next State} & \textbf{Description} & \textbf{Threats} \\
        \hline
        Idle    & -                        & Idle    & Drone is powered on.                                 & - \\			
        Idle    & Arm command received     & Arm     & Drone is armed and ready for takeoff.                & M, R, D \\
        Idle    & Error detected           & Error   & Drone has encountered an error.                      & M, R, D \\
        Arm     & Takeoff command received & Takeoff & Drone starts the takeoff process.                    & M, R, D \\
        Arm     & Disarm command received  & Disarm  & Drone is disarmed.                                   & M, R, D \\
        Arm     & Error detected           & Error   & Drone has encountered an error.                      & M, R, D \\
        Takeoff & Takeoff complete         & Mission & Drone is ready to perform its designated mission.    & S, M, R, D \\
        Takeoff & Error detected           & Error   & Drone has encountered an error during takeoff.       & S, M, R, D \\
        Mission & Mission complete         & Land    & Drone starts the landing process.                    & S, M, R, D \\
        Mission & Error detected           & Error   & Drone has encountered an error during the mission.       & S, M, R, D \\
        Land    & Landing complete         & Disarm  & Drone is disarmed after landing.                     & S, M, R, D \\
        Land    & Error detected           & Error   & Drone has encountered an error during landing.       & S, M, R, D \\
        Disarm  & Disarm complete          & Idle    & Drone is ready to be powered off.                    & M, R, D \\
        Disarm  & Error detected           & Error   & Drone has encountered an error after disarming.      & M, R, D \\
        Error   & Error resolved           & Idle    & Drone returns to Idle after resolving the error.     & S, M, R, D \\
        Error   & Error resolved           & Mission & Drone returns to Mission after resolving the error.  & S, M, R, D \\
        Error   & Critical error           & Land    & Drone attempts to land safely due to critical error. & S, M, R, D \\
        \hline
    \end{tabular}
    \\[1ex] 
    \footnotesize{\underline{\textbf{S}: Sensor Spoofing,  \textbf{M}: MITM, \textbf{R}: Replay, \textbf{D}: DoS/DDoS}}		
\end{table*}

\subsection{Feature Extraction for Drone Operation Analysis}~\label{sec:features}
We leverage the state transitions to evaluate MAVLink command sequences and identify the operational states and expected behavior. As the sequences of the transitions and commands are necessary for correct reasoning, our proposed IDS utilizes them to effectively detect potential security threats and anomalies. As such, DUDE-IDS monitors the MAVLink commands and sensor/actuator readings, applies feature analysis, and employs RNN-based anomaly detection. 

Table~\ref{tab:IDSFeatures} shows our extracted features based on monitored MAVLink command sequences and operational states. The \textit{packet direction} shows the communication direction (from drone to GCS or vice versa or between drones). Deviations in \textit{message length} may reveal tampered or malicious payloads. Long messages could mean command injection attacks, while short ones might be due to disrupted traffic. Monitoring \textit{sequence numbers} is vital for detecting inconsistencies like missing or out-of-order commands, which can point to packet loss, replay attacks, or message tampering. \textit{Source IDs} verify the legitimacy of incoming commands. Anomalies, such as commands from unknown sources, may signal spoofing or unauthorized access.  
Commands should be correctly mapped to the specific components for the correctness and integrity of the drone operations.  
Anomalies in \textit{component IDs} might indicate misrouted commands or attacks targeting subsystems, compromising the capabilities of the drone. \textit{Message IDs} help identify command types (e.g., \textit{mav\_cmd\_nav\_takeoff}, \textit{mav\_cmd\_nav\_land}, \textit{mav\_cmd\_nav\_waypoint}) and anomalies in these features can indicate attempts to manipulate the behavior of the drone, posing risks to its mission. \textit{Compatibility flags} ensure protocol adherence and prevent misuse (e.g., FC discards packets if flags have uncommon values). Such anomalies can indicate protocol violations or exploitation attempts, affecting the drone's functionality. 

\begin{table}[hbtp]
	\caption{DUDE-IDS Feature Set.}
	\label{tab:IDSFeatures}
	\centering
	\begin{tabular}{|p{3cm}|p{8cm}|}
			\hline
			\textbf{Feature Name} & \textbf{Description}\\
			\hline
			Packet direction       & Direction of the MAVLink packet.\\ \hline
			Message length         & Message length of the MAVLink command.\\ \hline
			Sequence number        & Sequence number of the MAVLink command.\\ \hline
			Source ID              & Source ID of the MAVLink command.\\ \hline
			Component ID           & Component ID of the MAVLink command.\\ \hline
			Message ID & Message ID of the MAVLink command.\\ \hline
			Compat/Incompat flags  & Flags that need to be recognized/ignored for the MAVLink compatibility.\\ \hline
			Current State          & Current state of the drone.\\ \hline
			Previous State         & Previous state of the drone.\\ \hline
			MAVLink command attributes & Values of each corresponding MAVLink command attribute.\\
			\hline
		\end{tabular}
\end{table}

The \textit{current state} and the \textit{previous state} of the drone provide contextual state information (based on Section~\ref{sec:states}). Unexpected state changes, like a sudden switch from \textit{Mission} to \textit{Idle}, might indicate unauthorized interventions or malfunctions. For example, during the \textit{Takeoff}, a drone is expected to ascend to a safe altitude before starting the \textit{Mission} state; however, an unexpected \textit{Land} command during the \textit{Takeoff} can signify an abnormal command injection or spoofing attempt. 

Each \textit{MAVLink command} has unique attributes. 
For example,\textit{ command\_long (1,0,22,0,0,0,0,0,0,0,20)} instructs the drone (\textit{target\_system = 1}) to take off (\textit{mav\_cmd\_nav\_takeoff}, using \textit{command ID 22}) to an altitude of 20 meters (\textit{param7 = 20}). The latitude and longitude parameters (\textit{param5} and \textit{param6}) set to \textit{0} means the drone should take off from its current location and not move laterally. 
Anomalies in these, such as unexpected parameter values, can indicate tampering or malicious intent, posing operational security risks. 

These features provide a comprehensive analysis input for our anomaly detection by allowing us to create a broad spectrum of operational and communication parameters and their interactions. The proposed DUDE-IDS for control can effectively model the normal behavior of the drone and identify deviations.

\subsection{Feature Encoding and Sequence Construction}\label{sec:encoding}
A key design choice in DUDE-IDS is to convert heterogeneous MAVLink and state-machine features into a unified time-ordered tensor suitable for LSTM inference. We define one \textit{timestep} as one observed MAVLink message on the MC-FC telemetry channel. For each message at time index $t$, we construct a feature vector $x_t \in \mathbb{R}^{d}$ by concatenating (i) protocol/header features, (ii) state-machine context, and (iii) command attributes, as follows.

\textbf{(i) Protocol/header features.}
We encode the packet direction as a binary value. Message length and sequence number are encoded as scalar values and z-score normalized. Source ID, component ID, and message ID are categorical identifiers; we encode them using one-hot encoding over the finite set observed in training. Compatibility/incompatibility flags are encoded as a fixed-width bit-vector. 
Table~\ref{tab:encoding_example} illustrates an example of how MAVLink protocol features are transformed from their original representation into the encoded numerical vector used by DUDE-IDS. 
After encoding, each MAVLink message is represented as a fixed-dimensional vector $x_t$, which is appended to the sliding temporal window before being processed by the LSTM network.

\begin{table}[hbtp]
\caption{Example encoding of MAVLink protocol/header features.}
\label{tab:encoding_example}
\centering
\small
\begin{tabular}{|l|l|l|}
\hline
\textbf{Feature} & \textbf{Original Value} & \textbf{Encoded Representation} \\
\hline
Packet direction & GCS $\rightarrow$ Drone & 1 \\
Message length   & 33 bytes                & 0.42 (z-score normalized) \\
Sequence number  & 145                     & 0.37 (z-score normalized) \\
Source ID        & 1                       & [1,0,0,0] (one-hot) \\
Component ID     & 190                     & [0,1,0] (one-hot) \\
Message ID       & MAV\_CMD\_NAV\_TAKEOFF & [0,0,1,0,0] (one-hot) \\
Compat flags     & 00000001                & [0,0,0,0,0,0,0,1] \\
\hline
\end{tabular}
\end{table}

\textbf{(ii) State-machine context.}
The current and previous operational states (Section~\ref{sec:states}) are encoded as one-hot vectors. This provides a coarse operational context that allows the temporal model to learn valid command/state co-occurrences (e.g., \textit{Takeoff} is expected after \textit{Arm}).

\textbf{(iii) Command attributes.}
MAVLink \texttt{COMMAND\_LONG} message parameters (\texttt{param1 $\ldots$ param7}) are included as normalized scalar values. 
Each parameter is scaled using z-score normalization based on the statistics of the training dataset:

\begin{equation}
x_{norm} = \frac{x - \mu}{\sigma}
\end{equation}
where $x$ is the original parameter value, $\mu$ is the mean, and $\sigma$ is the standard deviation computed from the training data. 
This normalization ensures that parameters with different physical ranges (e.g., altitude, velocity, or coordinates) contribute comparably during model training. 
For message types without certain parameters, the corresponding entries are set to zero and an additional mask bit is used to indicate missing values. 
This strategy preserves a fixed-dimensional representation across heterogeneous MAVLink message types. 
Table~\ref{tab:cmd_encoding_example} provides an example of encoding MAVLink command attributes. 
For example, a MAVLink \texttt{COMMAND\_LONG} message corresponding to a takeoff command with a target altitude of 20 meters would be encoded as shown in Table~\ref{tab:cmd_encoding_example}.

\begin{table}[hbtp]
\caption{Example encoding of MAVLink command attributes.}
\label{tab:cmd_encoding_example}
\centering
\small
\begin{tabular}{|l|l|l|}
\hline
\textbf{Parameter} & \textbf{Original Value} & \textbf{Encoded Value} \\
\hline
param1 (pitch)      & 0                     & 0.00 \\
param2 (unused)     & 0                     & 0.00 \\
param3 (unused)     & 0                     & 0.00 \\
param4 (yaw)        & 0                     & 0.00 \\
param5 (latitude)   & 0                     & 0.00 \\
param6 (longitude)  & 0                     & 0.00 \\
param7 (altitude)   & 20 m                  & 0.61 (normalized) \\
\hline
\end{tabular}
\end{table}

\textbf{Windowing and inference granularity.}
DUDE-IDS performs inference on a sliding window of length $L$ timesteps. Let $X_t = [x_{t-L+1},\dots,x_t] \in \mathbb{R}^{L \times d}$ denote the window ending at time $t$. The LSTM outputs a class prediction $\hat{y}_t$ for the entire window (many-to-one) corresponding to \{Normal, GPS Spoofing, MITM, Replay, DoS\}. We use a stride $s$ (overlap of $L-s$) to balance detection latency and computational cost. In our implementation, windows are created online as MAVLink messages arrive; if fewer than $L$ messages have been observed (e.g., immediately after boot), we left-pad with zeros and set the mask bits accordingly.

\subsection{LSTM-based Anomaly Detection for DUDE-IDS}

\begin{figure}[!h]
	\centering
	\includegraphics[width=.65\columnwidth]{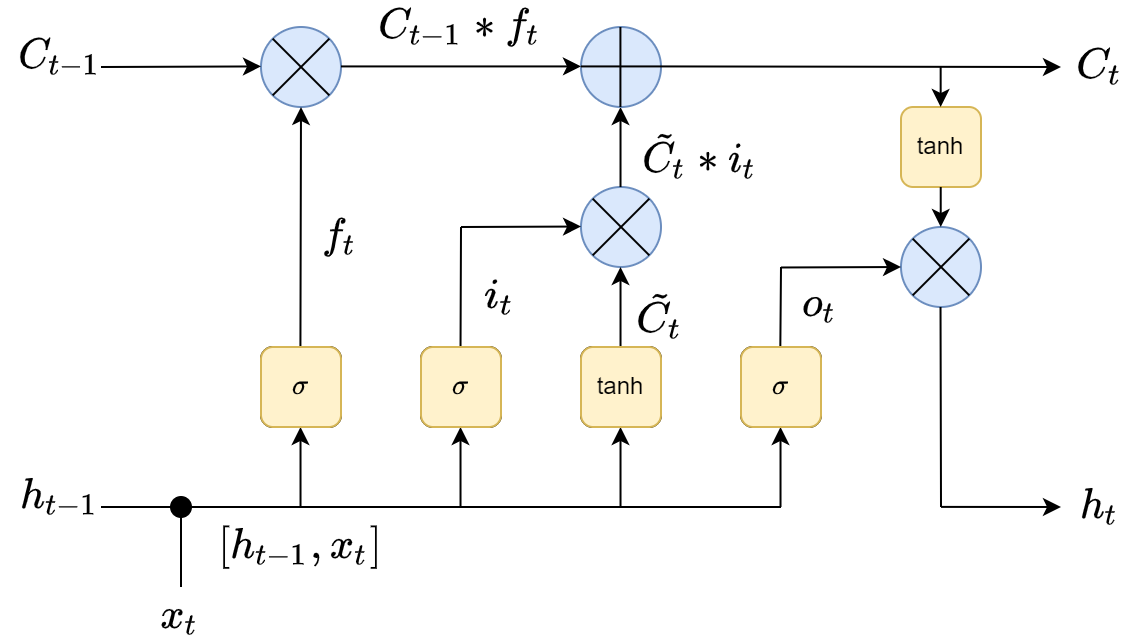}
	\caption{Architecture of LSTM~\cite{lstm:architecture}.}
	\label{fig:LSTM}
    \Description{Architecture of LSTM.}
\end{figure}

RNNs are a class of artificial NNs designed specifically for sequential data processing~\cite{rnn:medsker}. The standard RNN architecture~\cite{rnn:Schuster} provides a straightforward approach to modeling sequential data, but it has limitations in handling long-term dependencies, and it has a probability of vanishing gradient issues, which can reduce its effectiveness in detecting complex anomalies during drone operations. In contrast, Long Short-Term Memory (LSTM)~\cite{lstm:Hochreiter} networks outperform standard RNNs due to their gating mechanisms and utilizing/maintaining information over long sequences. In addition to remembering/utilizing information from previous time sequences over a long period, LSTM also shows improved capabilities to filter out noise in the data and focus on relevant temporal patterns, which make LSTM highly effective for identifying indirect and gradual anomalies in drone sensor data and command sequences. Using these features, this work leverages LSTM networks for anomaly detection in drone operations (based on the features given in Section~\ref{sec:features}). The LSTM architecture is shown in Figure~\ref{fig:LSTM} of which functions are explained in Equations~\ref{eq:it} to~\ref{eq:ht}, with the details of how it can be leveraged for the drone operation anomalies.

\begin{equation}~\label{eq:it}
	i_t = \sigma(W_i \cdot [h_{t-1}, x_t] + b_i)
\end{equation}
\begin{equation}~\label{eq:tCt}
	\tilde{C}_t = \tanh(W_C \cdot [h_{t-1}, x_t] + b_C)
\end{equation}
\begin{equation}~\label{eq:ft}
	f_t = \sigma(W_f \cdot [h_{t-1}, x_t] + b_f)
\end{equation}
\begin{equation}~\label{eq:Ct}
	C_t = f_t * C_{t-1} + i_t * \tilde{C}_t
\end{equation}
\begin{equation}~\label{eq:ot}
	o_t = \sigma(W_o \cdot [h_{t-1}, x_t] + b_o)
\end{equation}
\begin{equation}~\label{eq:ht}
	h_t = o_t * \tanh(C_t)
\end{equation}

The cell state in LSTM is used to capture and retain long-term dependencies in sequential data~\cite{lstm:architecture}. 
The non-linear activation functions play a critical role in this process: the sigmoid function ($\sigma$) maps values to the range $[0,1]$, effectively acting as a gating mechanism to control information flow, while the hyperbolic tangent function ($\tanh$) scales values to $[-1,1]$, enabling stable representation of both positive and negative correlations in the data.

\textit{The input gate} ($i_t$) and the candidate cell state ($\tilde{C}_t$) in Equations~\ref{eq:it} and~\ref{eq:tCt} regulate how new observations are incorporated into the memory. Specifically, $i_t$ determines the importance of incoming information, while $\tilde{C}_t$ represents the candidate content to be added. This mechanism allows the LSTM to selectively integrate meaningful changes in the system, such as sudden deviations in altitude, unexpected MAVLink command patterns, or abnormal sensor correlations. For example, during a replay or spoofing attack, inconsistencies between expected and observed command sequences can be captured through this selective update process.

\textit{The forget gate} ($f_t$) in Equation~\ref{eq:ft} controls how much of the previous cell state ($C_{t-1}$) should be retained. By assigning values close to 0 or 1, it effectively filters out irrelevant or outdated information while preserving critical temporal context. This is particularly important in drone operations, where transient noise (e.g., sensor jitter or environmental disturbances such as wind gusts) should not trigger false alarms, whereas persistent deviations over time may indicate malicious activity. Thus, the forget gate enables the model to distinguish between short-term fluctuations and sustained anomalous behavior.

Together, these gating mechanisms allow the LSTM to maintain a robust temporal representation of drone operations, capturing both immediate anomalies and long-term behavioral deviations across sensor data and command sequences.

The cell state update in Equation~\ref{eq:Ct} combines the effects of the forget and input gates to maintain a running context of the drone's operational state, which is essential for understanding long-term dependencies, such as gradual deviations in command patterns or sensor readings that build up to an attack scenario. 
As illustrated in Figure~\ref{fig:LSTM}, the \textit{forget gate} ($f_t$) controls the retention of prior information, while the \textit{input gate} ($i_t$) regulates the incorporation of new observations into the cell state. 

\textit{The output gate} ($o_t$) and the hidden state ($h_t$) in Equations~\ref{eq:ot}~and~\ref{eq:ht} use this updated cell state to determine the next predicted state, effectively capturing deviations from normal operational patterns. 
For example, a hidden state $h_t$ that reflects an unexpected command sequence (e.g., a \textit{Land} command while the drone is in the middle of a \textit{Mission} state) can be indicative of command injection attacks.

Figure~\ref{fig:dude_lstm_arch} shows the final DUDE-IDS pipeline and the employed many-to-one LSTM classifier. The input is a sliding window tensor $X_t \in \mathbb{R}^{L \times d}$ constructed as described in Section~\ref{sec:encoding}. The LSTM stack produces the final hidden state, which is passed through a fully-connected layer and softmax to output the predicted class for the window.

\begin{figure}[htbp]
    \centering
    \label{fig:dude_lstm_arch}
    \includegraphics[width=0.95\linewidth]{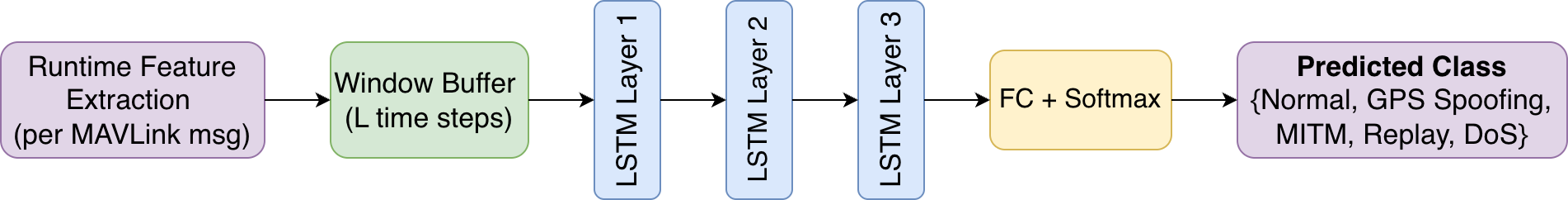}
    \caption{DUDE-IDS LSTM architecture.}
    \label{fig:dude_lstm_arch}
\end{figure}

\section{Experimental Setup} \label{sec:ExperimentalSetup}

\subsection{Experimental Testbed} 
For experimentation, implementation, and evaluation, we custom-built our testbed drone using a Raspberry Pi 4 Model B for the MC and Pixhawk 2.4.8 for the FC, as they are commonly used pairs for drone operations. We also observe that some studies use NVIDIA Jetson Nano due to its GPU accelerators for drone AI operations (preferred for tasks such as object tracking and collision avoidance). For our experiments, we utilized a 3DR Radio (to facilitate Radio Frequency (RF) between a drone MC and GCS), a HackRF One (for capturing, analyzing, and transmitting radio signals), and the Universal Radio Hacker (URH) software (to analyze radio signals and create custom signal profiles for GPS spoofing, MITM, DoS, and replay attacks)~\cite{tools:Johannes} as well as GPS-SDR-SIM simulator (to generate fake GPS signals)~\cite{tools:gpssdr}.

\subsection{Dataset Generation}
The dataset used to detect anomalies in MAVLink commands and sensor/actuator readings was generated by simulating a series of flight scenarios under both normal and cyberattack conditions. 

\subsubsection{Dataset Summary and Labeling}\label{sec:dataset_summary}
To enable reproducibility and to clarify the scope of the contributed dataset, Table~\ref{tab:dataset_summary} summarizes the generated data. Each recorded MAVLink message is timestamped and labeled according to the scenario being executed at that time (Normal, GPS Spoofing, MITM, Replay, or DoS). Labels are assigned at the \textit{scenario interval} level and inherited by the windows $X_t$ extracted from those intervals (Section~\ref{sec:encoding}). Windows, which span multiple labels, are discarded to avoid ambiguous supervision.

\begin{table}[htbp]
\caption{DUDE-IDS dataset summary (DIY drone testbed).}
\label{tab:dataset_summary}
\centering
\small
\begin{tabular}{|l|c|}
\hline
\textbf{Item} & \textbf{Value} \\
\hline
\# flight scenarios (Normal + attacks) & 5 \\
Runs per scenario & 10 \\
Total flight time (minutes) & $\approx 100$--$120$ \\
Total MAVLink messages logged & $\approx 500{,}000$ \\
Classes & Normal, GPS Spoofing, MITM, Replay, DoS \\
Window length ($L$) / stride ($s$) & $20 / 5$ \\
Train/test split & 80\% / 20\% \\
Standardization & z-score (train statistics) \\
\hline
\end{tabular}
\end{table}

The dataset with (i) raw MAVLink logs, (ii) synchronized sensor streams, (iii) scenario timestamps, and (iv) scripts to reproduce window extraction and feature encoding will be shared with the research community on our group website, as such a dataset does not exist in the literature.

\subsubsection{Monitoring Module Setup}
We created and deployed a \textit{real-time monitoring module} to continuously log telemetry data, including sensor readings (e.g., IMU, GPS, barometer), actuator commands (e.g., motor speeds), and all MAVLink commands sent to our custom-built drone. The monitoring system was implemented as an onboard process (written in Python) with minimal overhead to ensure that the drone's operational performance remained unaffected. 

\subsubsection{Flight Scenarios}
For training, fine-tuning, and testing, we created multiple flight scenarios and monitored drone operations under both normal operational conditions and cyberattacks. In our \textit{baseline} scenario, MC executed the basic operations under normal conditions only. For this case, the MC may be used for object detection by utilizing the YOLO V8 model~\cite{tools:yolov8}. We also deployed the proposed DUDE-IDS with no load (as \textit{idle}). During \textit{normal operations}, the drone followed pre-defined flight paths without interference, capturing baseline data for sensor and actuator readings as well as MAVLink commands.

The cyberattack scenarios were designed to represent realistic operational environments under malicious interference, exploiting specific vulnerabilities in the communication and control systems of the flight controller. In the \textit{GPS spoofing attack} scenario, we manipulated the GPS data in real-time, leading to anomalies in both sensor readings and actuator responses. We transmitted fake GPS signals based on the daily GPS broadcast ephemeris file (\textit{brdc})~\cite{tools:nasa} to the GPS module of the Pixhawk-based FC to manipulate its perceived location. The GPS-SDR-SIM tool was used to transmit fake GPS signals using HackRF One. 
For the \textit{replay attack} scenario, previously` recorded sensor and command data were sent/replayed to the drone during a different mission, resulting in detectable inconsistencies. The URH software facilitated the capture and retransmission of these signals, allowing for the testing of the ability of the DUDE-IDS to detect malicious commands. 
In the \textit{DoS attack} scenario, using the URH software and the 3DR radio, a high-volume traffic load (sends a mixture of commands at high frequency) was sent to the drone, causing delayed and/or dropped commands, with the resulting anomalies in the drone’s behavior to evaluate the DUDE-IDS resilience to communication overload and its ability to detect malicious packets and protect operational integrity under such conditions. Finally, during the \textit{MITM attack}, the specific command sequences (e.g., incorrect latitude and longitude, and incorrect waypoint commands) were sent to the drone in mid-flight, causing deviations from the drone's expected behavior.

Each flight scenario was executed ten times to account for sensor noise variability and environmental factors. In each flight scenario, data was labeled as normal operation or malicious accordingly.

\subsection{LSTM and CNN Model Generation}
After the generation of the dataset and extraction of the relevant features (given in Table~\ref{tab:IDSFeatures}), we standardized the features to have a mean of zero and a standard deviation of one. This standardization process ensures that the input data is normalized, which is crucial for the model performance. 
The data is then split into training (80\%) and testing (20\%) sets to facilitate model evaluation. 

Our LSTM-based RNN model consists of an input layer, multiple LSTM layers to learn temporal patterns, and a fully connected output layer that maps the learned representations to the output classes. \textit{Hyperparameters} such as the number of hidden units (50), the number of layers (3), and the learning rate (0.001) are carefully selected to optimize the model's performance \textit{empirically}, as summarized in Table~\ref{tab:lstm_hparams}. 

\begin{table}[htbp]
\caption{DUDE-IDS LSTM hyperparameters.}
\label{tab:lstm_hparams}
\centering
\small
\begin{tabular}{|l|c|}
\hline
\textbf{Hyperparameter} & \textbf{Value} \\
\hline
Window length ($L$) & 20 \\
Stride ($s$) & 5 \\
Input dimension ($d$) & $\approx 60$--$80$ \\
\# LSTM layers & 3 \\
Hidden units per layer & 50 \\
Dropout & 0.2 \\
Loss & Cross-entropy \\
Optimizer & Adam \\
Learning rate & 0.001 \\
Batch size & 32 \\
Epochs & 10 \\
\hline
\end{tabular}
\end{table}

We selected 50 hidden units to balance between model complexity and computational cost. This number is large enough to capture complex temporal patterns in the drone's operational data while minimizing the risk of overfitting (even though higher accuracy results may be observed temporarily) and keeping computational demands manageable. We opted for 3 LSTM layers, as this depth provided an optimal balance between model capacity and computational stability. Additional layers, while beneficial for learning deeper features, could introduce vanishing gradient issues. Our initial experiments showed that 3 layers were sufficient to capture long-term dependencies without negatively impacting performance. A learning rate of 0.001 was chosen for steady convergence during training. A higher learning rate of $lr = 0.01$ made the model learn very quickly, reducing the loss to near zero early on and giving perfect test accuracy (100\%) and F1-score (1.0000). However, this also means the model overfits and likely fails on new data. On the other hand, a lower learning rate of $lr = 0.0001$ caused the model to learn too slowly, with little improvement in loss and test accuracy of 80.77\%, along with an F1-score of 0.7218, showing the model did not learn effectively. Therefore, \textit{our empirical results} show that a learning rate of $lr = 0.001$ worked best, achieving a good test accuracy of 98.08\% and an F1-score of 0.9713.

We preferred the cross-entropy loss function for multi-class classification and the Adaptive Optimizer (Adam) for model weight updates based on the computed gradients. The training is conducted over multiple epochs ($epoch = 10$), where the model learns to minimize the loss function by adjusting its weights. In our experiments, using fewer epochs (such as 2 or 3) resulted in incomplete learning, as seen in the lower test accuracy (80.77\%) and F1-score (0.7218), indicating the model had not fully captured the underlying patterns in the data. On the other hand, when training for 4 epochs or more, the model achieved consistent and higher performance, with test accuracy reaching 98.08\% and F1-score stabilizing around 0.9713. During the training, mini-batches of data are fed into the model to enhance generalization and prevent overfitting. After the training, the evaluation is performed by calculating the accuracy, F1, precision, and recall scores.

\begin{figure}[htbp]
    \centering
    \includegraphics[width=1\linewidth]{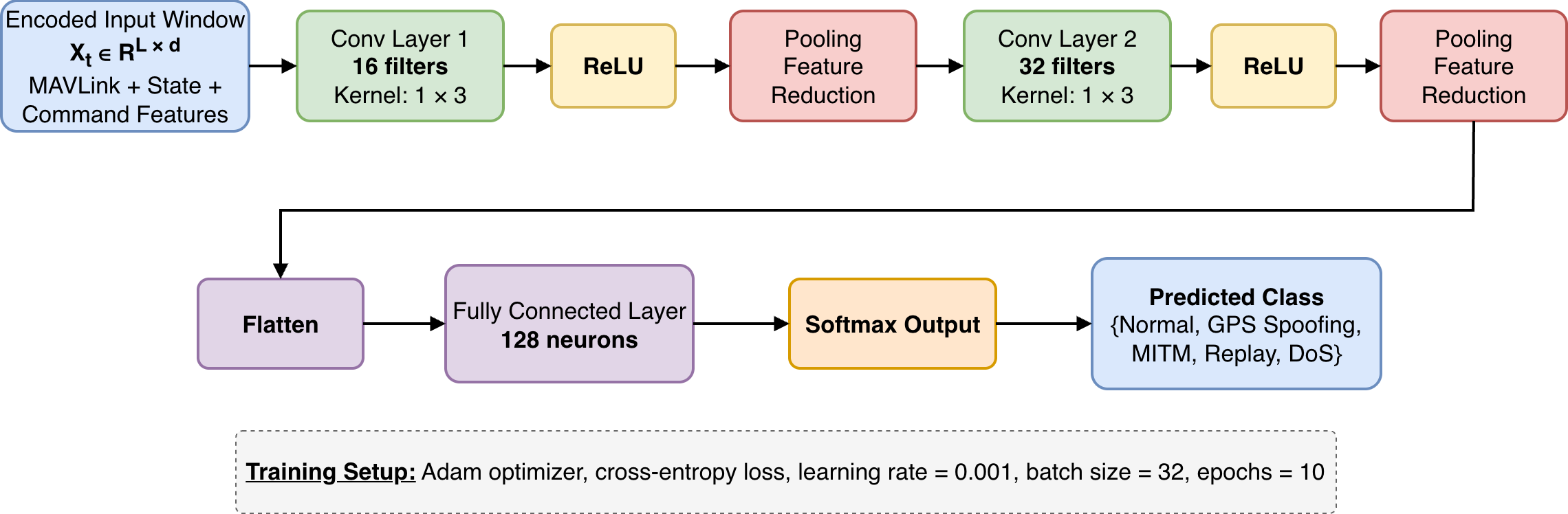}
    \caption{CNN-based DUDE-IDS implementation used for comparison.}
    \label{fig:CNN}
\end{figure}

For comparison purposes, we also implemented a CNN-based model as shown in Figure~\ref{fig:CNN}. Our CNN-based solution includes two convolutional layers (with 16 and 32 filters, respectively), followed by a fully connected layer of 128 neurons, and an output layer that maps learned features to the target classes. The small kernel size of 1×3 ensures that the model captures local patterns in the drone's operational data. The number of filters was selected to balance learning capacity and computational complexity. A learning rate of 0.001 was chosen empirically. The training was done using mini-batches with a batch size of 32 over 10 epochs, providing sufficient learning without over-complicating the model. For optimization, we used the Adam optimizer and the cross-entropy loss function, similar to the RNN model.

\section{Experimental Results} 
\label{sec:ResultsandDiscussions}

The proposed LSTM-based RNN and also CNN models were both evaluated using our test dataset mentioned in Section~\ref{sec:dataset_summary}. We first evaluate the accuracy of the two approaches when a Raspberry Pi is used as the MC (Table~\ref{tab:classification}). The LSTM model achieves a classification accuracy of 98.08\%, with a precision of 0.9620, recall of 0.9808, and an F1 score of 0.9713. The CNN model demonstrates less effective classification accuracy of 96.45\%, with a precision of 0.9302, a recall of 0.9645, and an F1 score of 0.9470. Therefore, we choose the LSTM-based RNN model over CNN due to its superior ability to capture temporal dependencies, which are crucial for maintaining operational security in drone-related tasks. 

\begin{table}[htbp]
    \centering
    \caption{Classification performance comparison of LSTM-based RNN and CNN models for DUDE-IDS (superior results are shown in bolder fonts).}
    \begin{tabular}{|c|c|c|} \hline
         \textbf{Parameter} & \textbf{LSTM} & \textbf{CNN} \\ \hline\hline
         \textbf{Accuracy} & \textbf{98.08}\% & 96.45\% \\ \hline
         \textbf{Precision} & \textbf{0.9620} & 0.9302 \\ \hline
         \textbf{Recall} & \textbf{0.9808} & 0.9645 \\ \hline
         \textbf{F1 score} & \textbf{0.9713} & 0.9470 \\ \hline
    \end{tabular}
    \label{tab:classification}
\end{table}

In addition to classification accuracy, we evaluate the decision time of both models to assess their suitability for real-time deployment in Table \ref{tab:latency}. Specifically, we measure the average inference time per input window ($T_{\text{inf}}$) on the Jetson Nano-based MC. 

The LSTM-based model achieves an average inference time of below \textit{25 ms} per window, while the CNN-based model requires approximately \textit{12 ms} per window. Given a window size of $L = \textit{20}$ messages and a streaming input rate, the effective detection latency is bounded by the time required to collect the window plus the inference time. 

Our results indicate that the overall detection delay remains within \textit{250--300 ms}, demonstrating that the proposed DUDE-IDS can operate in real-time or near real-time on resource-constrained drone platforms. Although the CNN-based model exhibits lower inference latency due to its parallelizable architecture, the LSTM-based model achieves higher detection accuracy by effectively modeling temporal dependencies across sequential MAVLink messages. This trade-off highlights the suitability of LSTM for security-critical drone applications where capturing temporal anomalies is essential.

\begin{table}[htbp]
\centering
\caption{Approximate inference latency comparison on Raspberry Pi-based MC.}
\label{tab:latency}
\begin{tabular}{|c|c|c|}
\hline
Model & Inference Time & Estimated Detection Delay \\
\hline
LSTM & $<25$ ms/window & 225--275 ms \\
CNN  & $\approx 12$ ms/window & 212--262 ms \\
\hline
\end{tabular}
\end{table}

For a real-time scenario, the power consumption and resource requirements are also critical because drones have limited computational capabilities and limited batteries. Hence, we also compare our proposed DUDE-IDS in terms of power consumption and resource usage. The power consumption statistics of two MC systems (Raspberry Pi and Jetson Nano) under various operational conditions are summarized in Table~\ref{table:rpi_power_stats} and Table~\ref{table:nano_power_stats}, respectively. These conditions include a baseline state where no additional processes are running, running of an object detection model (we used YOLO V8), an IDLE state where just DUDE-IDS is running, and various types of cyberattacks when DUDE-IDS is running. It provides a comprehensive overview of how different activities affect energy usage. We performed 10 iterations for each condition and presented the average values in the table. The statistics reported are the minimum (Min), mean (Mean), maximum (Max), and standard deviation (Std) of power consumption in watts (W).

\begin{figure}[htbp]
	\centering
	\includegraphics[width=.85\columnwidth,]{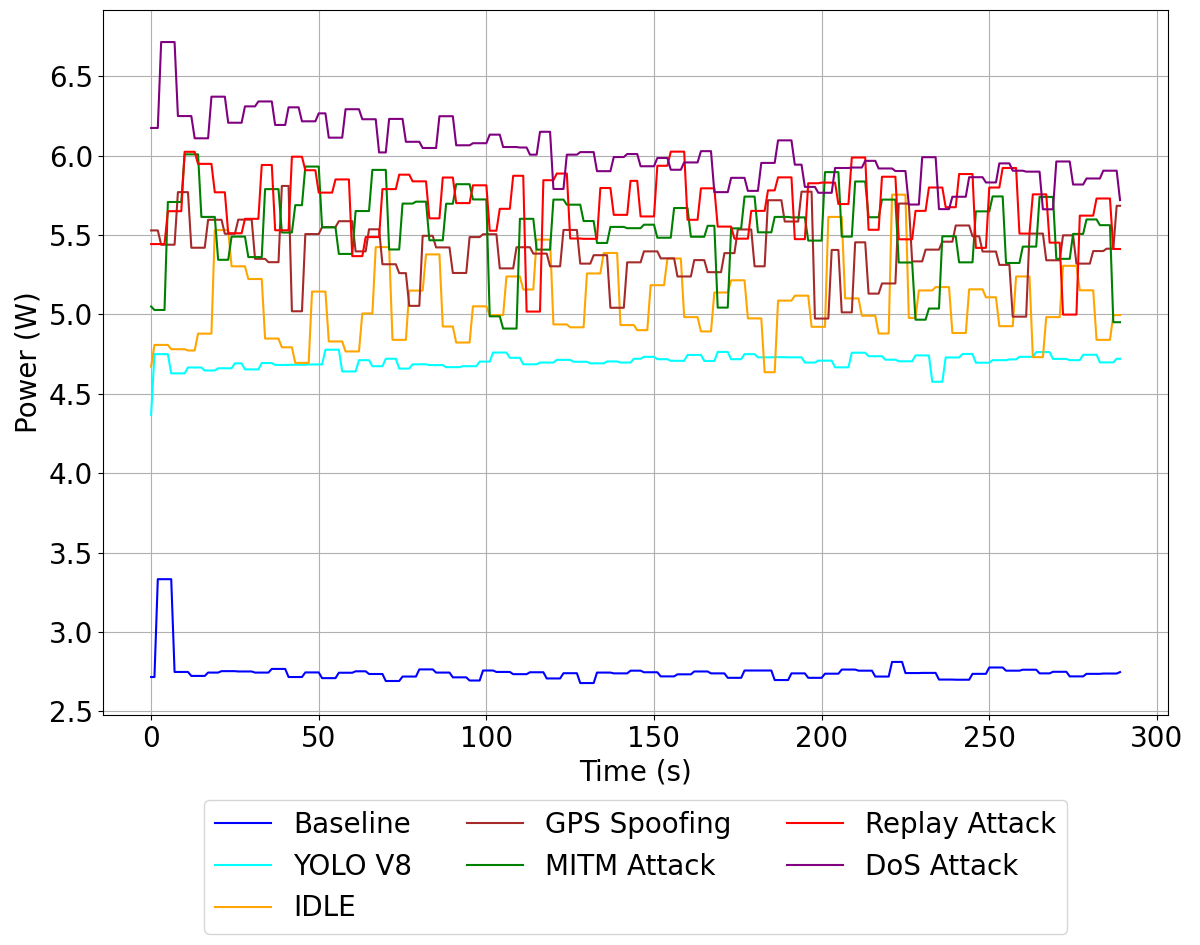}
	\caption{Power consumption over time for the Raspberry Pi-based MC for the test drone running IDLE, baseline, YOLOv8, and DUDE-IDS operations.}
	\label{fig:power_vs_time_rpi}
    \Description{Power consumption over time for the Raspberry Pi-based MC.}
\end{figure}

Table~\ref{table:rpi_power_stats} provides insights into the power consumption of the Raspberry Pi under various conditions, including normal operation, YOLO V8, and several types of cyberattacks. Figure~\ref{fig:power_vs_time_rpi} demonstrates the power consumption (in Watts) of these scenarios over time for detailed analysis. In the baseline condition, the power consumption remains low and stable at around 2.75 W, indicating minimal system activity. However, when the YOLO V8 model is executed, the power consumption jumps to approximately 4.7 W,  reflecting the additional computational load. In the IDLE state, the power usage rises further to about 5.3 W with more variability. When subjected to cyberattacks, such as MITM, Replay, and GPS spoofing, the power consumption fluctuates between 5.5 W and 6.0 W. This indicates increased resource utilization by the system during these attacks. Among these, the DoS attack stands out with the highest power consumption, averaging around 6.0 W and spiking to 6.5 W, showcasing its resource-intensive nature. During the experiments, we noticed that the DUDE-IDS latency was negligible; however, 3.5 out of the 4 Raspberry Pi cores were utilized during execution. This was expected since the Raspberry Pi lacks an additional ML processor, which prompted us to test with the Jetson Nano.

\begin{table}[htbp]
	\caption{Raspberry Pi-based MC power consumption.}
	\label{table:rpi_power_stats}
	\centering
	\begin{tabular}{|l|c|c|c|c|}
		\hline
		\textbf{Condition}  & \textbf{Min (W)} & \textbf{Mean (W)} & \textbf{Max (W)} & \textbf{Std (W)} \\ \hline
		Baseline            & 2.68             & 2.75              & 3.33             & 0.08\\ \hline
		YOLO V8             & 4.37             & 4.70              & 4.78             & 0.04\\ \hline
		IDLE                & 4.64             & 5.07              & 5.75             & 0.24\\ \hline
		GPS Spoofing        & 4.97             & 5.40              & 5.81             & 0.18\\ \hline
		MITM Attack         & 4.91             & 5.53              & 6.01             & 0.24\\ \hline
		Replay Attack       & 5.00             & 5.69              & 6.03             & 0.22\\ \hline
		DoS Attack          & 5.66             & 6.02              & 6.71             & 0.20\\ \hline
	\end{tabular}
\end{table}

\begin{figure}[hbtp]
	\centering
	\includegraphics[width=.85\columnwidth]{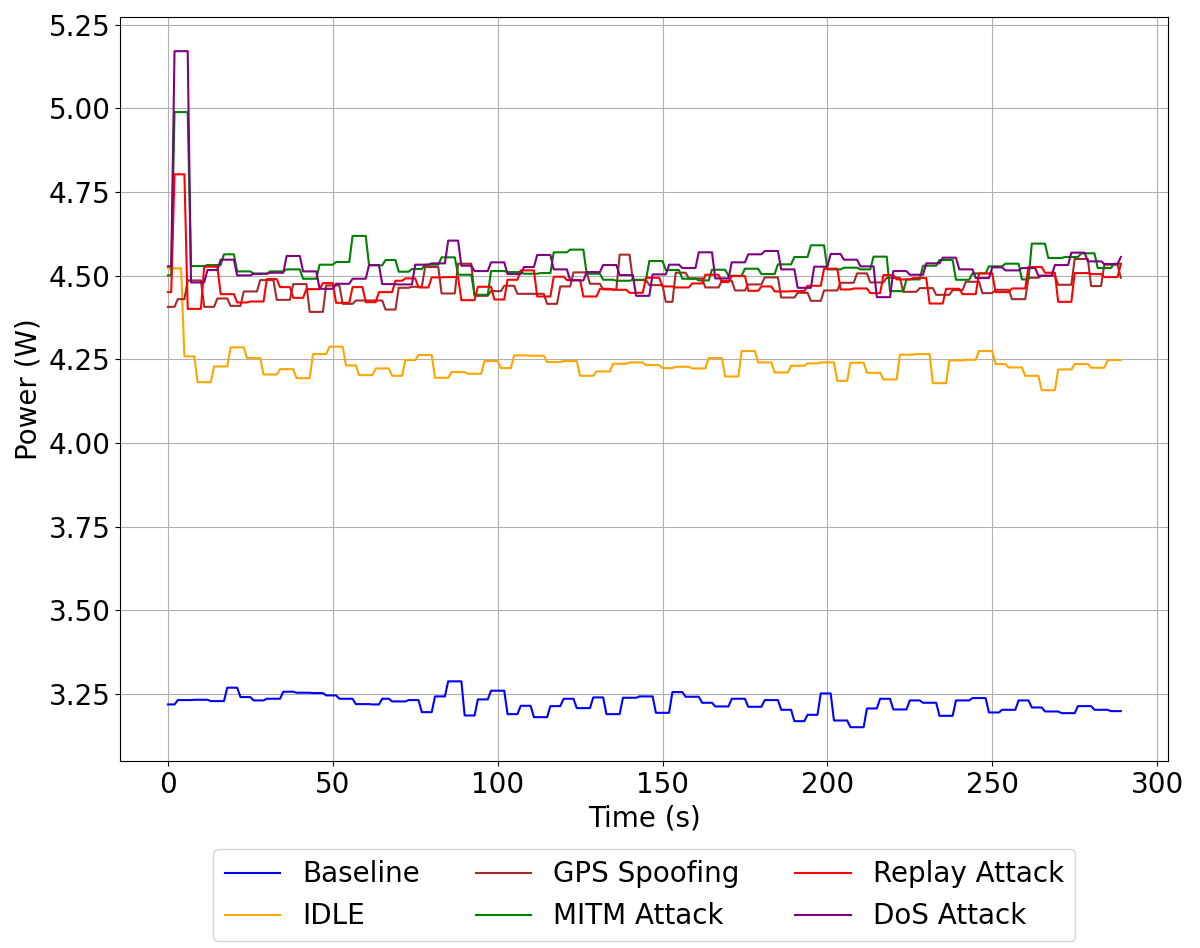}
	\caption{Power consumption over time for the Jetson Nano-based MC for the test drone running IDLE, baseline, YOLOv8, and DUDE-IDS operations.}
	\label{fig:power_vs_time_nano}
    \Description{Power consumption over time for the Jetson Nano-based MC.}
\end{figure}

Table~\ref{table:nano_power_stats}, along with Figure~\ref{fig:power_vs_time_nano}, provides insights into the power consumption of the Jetson Nano under normal operation and cyberattacks, similarly. In the baseline condition, the Jetson Nano demonstrates low and stable power consumption at around 3.22 W, indicating minimal system requirements. In the IDLE state, the power usage increases to approximately 4.54 W, with slight oscillations. During cyberattacks such as MITM, Replay, and GPS Spoofing, the power consumption rises further, fluctuating between 4.50 W and 4.75 W, reflecting the increased demand on system resources as it processes data. Among these, the DoS attack stands out with an average power consumption of 4.53 W, spiking as high as 5.25 W. We also observed that only 1 out of the 4 cores in Jetson Nano were utilized during DUDE-IDS execution, as the Jetson Nano is more capable for ML/AI algorithms. These results show that the Jetson Nano handles these attacks with less variability, highlighting the Jetson Nano-based system's capacity for handling more complex tasks efficiently due to its dedicated GPU.

\begin{table}[htbp]
	\caption{Jetson Nano-based MC power consumption.}
	\label{table:nano_power_stats}
	\centering
	\begin{tabular}{|l|c|c|c|c|}
		\hline
		\textbf{Condition} & \textbf{Min (W)} & \textbf{Mean (W)} & \textbf{Max (W)} & \textbf{Std (W)} \\ \hline
		Baseline           & 3.15             & 3.22              & 3.29             & 0.03\\ \hline
		IDLE               & 4.16             & 4.24              & 4.52             & 0.05\\ \hline
		GPS Spoofing       & 4.39             & 4.47              & 4.56             & 0.04\\ \hline
		MITM Attack        & 4.44             & 4.53              & 4.99             & 0.07\\ \hline
		Replay Attack      & 4.40             & 4.47              & 4.80             & 0.05\\ \hline
		DoS Attack         & 4.44             & 4.53              & 5.17             & 0.09\\ \hline
	\end{tabular}
\end{table}

The LSTM-based DUDE-IDS exhibited a superior classification accuracy with lower false positive and false negative rates, and effective detection of cyberattacks. However, this came at the cost of increased power consumption and CPU core usage, particularly on the Raspberry Pi, where the system utilized 3.5 out of 4 cores during operation. Additionally, we recognize that factors such as sensor noise, environmental variability, and evolving attack patterns in real-world scenarios may increase the risk of false detections. These risks can be mitigated by threshold adjustment and ensemble methods, as a future study to enhance the robustness and reliability of drone security.

In summary, the LSTM-based DUDE-IDS system demonstrated its applicability for real-time or near real-time operation on drones. Our experiments validated its effectiveness in detecting various types of cyberattacks by utilizing Raspberry Pi and Jetson Nano. Despite the computational constraints of drones, the system achieved a classification accuracy of 98.08\% while maintaining minimal latency. This shows that DUDE-IDS is capable of real-time anomaly detection and intrusion prevention by providing a practical solution for the operational security of drones. Since our solution relies on analyzing MAVLink commands and sensor data from a DIY drone, \textit{it is adaptable to a wide range of drones that use MAVLink}, which highlights its scalability to more advanced or commercial drone platforms. The dataset will be available on our research group website~\cite{smartcyberspaceSmartCyber}.

In comparison, \textit{transformer-based models}, such as lightweight variations of the Vision Transformer (ViT) or Linformer~\cite{jin2025linformer, wang2020linformer}, have demonstrated improvements in parallelization due to their self-attention mechanisms. These models can reduce inference times by eliminating the sequential data dependency inherent to LSTMs. However, their application in drone scenarios is limited by factors such as memory overhead, since transformers often require substantial hardware acceleration (e.g., GPUs or TPUs), which many drones lack~\cite{huang2022hardware, zhong2023transformer}. 
While our Jetson Nano-based solution could potentially support these transformer models, this may not be a generic solution for all drone platforms, especially those with constrained resources. Furthermore, real-time evaluation and optimization for power efficiency would be necessary to ensure applicability across diverse platforms.
Another alternative approach is model pruning, which aims to reduce the complexity of neural networks by removing redundant weights and neurons. Applying structured pruning techniques to our LSTM model could lower the power consumption and core usage, which is considered our future work.  

\subsection{Baselines and Runtime Evaluation}\label{sec:baselines_runtime}
To contextualize the benefit of temporal modeling, we compare DUDE-IDS against representative non-sequential baselines trained on the same encoded features aggregated over a window (mean/std/min/max) rather than fed as a sequence. We report accuracy, precision, recall, and F1 to remain consistent with Section~\ref{sec:ResultsandDiscussions}.

\subsubsection{Baselines} We include: (i) a rule-based state-transition validator that flags invalid state/command pairs using Table~\ref{tab:StateTransition}, (ii) Random Forest (RF) on window-aggregated features, and (iii) One-Class SVM / Isolation Forest trained on Normal-only data (anomaly detection setting). These baselines represent common IDS design points with lower computational complexity than LSTMs.

\subsubsection{Runtime and Throughput} In addition to power consumption (Tables~\ref{table:rpi_power_stats}--\ref{table:nano_power_stats}), we measure inference latency per window and sustained throughput. Let $T_{\text{inf}}$ denote the average inference time for one window and $R$ denote the maximum sustained processing rate in windows/s. 
These results directly quantify real-time feasibility: with window stride $s$, the effective detection delay is bounded by $\approx s$ messages plus $T_{\text{inf}}$, enabling near real-time alerting while preserving temporal context needed for replay cadence and MITM drift detection. 
Table~\ref{tab:runtime_stats} summarizes these measurements. 
From the results, we can observe that NVIDIA Jetson Nano provides a higher throughput and lower inference time per window. This is, in fact, anticipated due to Jetson Nano's high parallelism as it was designed for these tasks. 

\begin{table}[htbp]
\caption{Runtime statistics of DUDE-IDS inference (window-based).}
\label{tab:runtime_stats}
\centering
\small
\begin{tabular}{|l|c|c|}
\hline
\textbf{Platform} & $\mathbf{T_{\text{inf}}}$ \textbf{(ms/window)} & \textbf{Throughput $R$ (windows/s)} \\
\hline
Raspberry Pi 4 & $\approx 20$--$25$ & $\approx 40$--$50$ \\
NVIDIA Jetson Nano & $\approx 8$--$12$ & $\approx 80$--$120$ \\
\hline
\end{tabular}
\end{table}

\section{Further Discussions}
\label{sec:discussion}
\textbf{Scalability to diverse drone platforms.}
DUDE-IDS is designed using MAVLink message semantics and a platform-agnostic state abstraction (please refer to Section~\ref{sec:states}), rather than airframe-specific dynamics. Therefore, our approach is generalized for MAVLink-enabled drones and can be further extended by (i) re-enumerating observed IDs for one-hot/embedding tables, (ii) adjusting the state-machine mapping if flight modes differ, and (iii) re-normalizing sensor/command ranges during calibration flights. For commercial platforms that utilize MAVLink (or MAVLink-bridged telemetry), the same pipeline applies with minimal modification; for non-MAVLink systems, the method can be ported by defining an equivalent command vocabulary and state abstraction.
Even though our current study is performed using a custom/DIY drone, the drone has a commonly used FC and MC pair, and is based on a finite yet fundamental set of scenarios. While MAVLink-centric design supports portability, additional experiments on multiple airframes and mission profiles (e.g., different sensors, flight modes, and RF stacks) can be applied straightforwardly to fully quantify generalization.

\textbf{Environmental variability and sensor noise.}
Real-world missions are affected by environmental uncertainties such as wind gusts, multipath GPS errors, RF interference, and payload-dependent dynamics. Such uncertainties can shift feature distributions and increase false alarms. Our DUDE-IDS solution mitigates this by (i) including multi-sensor context (e.g., IMU/GPS/barometer consistency) in the feature vector, and (ii) using sequence modeling to distinguish transient noise from sustained deviations. As a practical deployment step, thresholds (or softmax confidence cutoffs) can be further calibrated per mission and environment profile, and periodic re-calibration flights can update normalization statistics. 

\textbf{Evolving attack patterns and adversarial adaptation.}
Attackers may vary replay cadence, blend benign commands with malicious ones, or attempt low-and-slow parameter drift. Since DUDE-IDS models temporal correlations across protocol fields, state context, and command attributes, these strategies remain detectable when they induce inconsistent cross-feature dynamics over time. To further harden the system, future work can incorporate drift detection (monitoring distributional change), incremental fine-tuning with recent benign mission logs, and lightweight ensembles (e.g., combining LSTM with a rule-based state validator) to reduce false positives without sacrificing detection sensitivity.

\section{Conclusion}\label{sec:Conclusion}

The security of drones is crucial due to their increasing deployment in sensitive applications such as surveillance, logistics, and disaster response, where cyberattacks can disrupt operations and pose significant safety risks. To address these concerns, we emphasized the importance of sequential data analysis, proposed various system states for anomaly detection, and designed an architecture tailored for analyzing MAVLink-based operations. This paper has demonstrated the critical role of RNNs in enhancing the security and reliability of drone operations through effective anomaly detection. By focusing on LSTM networks, we have shown that these models can accurately capture and analyze the sequential nature of drone sensor data and commands, identifying both simple and complex anomalies. The LSTM-based RNN model achieved a classification accuracy of 98.08\%, surpassing the CNN model's accuracy of 96.45\%. In addition, the LSTM model maintained higher precision (0.9620), recall (0.9808), and F1 score (0.9713) which makes it more suitable for DUDE-IDS.

Our experiments also evaluated the power consumption and resource utilization of Raspberry Pi and Jetson Nano. The Raspberry Pi showed significant power consumption under various attack scenarios, reaching a peak of 6.5 W during DoS attacks and utilizing 3.5 out of its 4 cores during execution. In contrast, the Jetson Nano demonstrated more stable and lower power consumption, with a peak of around 5.25 W, while only utilizing 1 out of its 4 cores. The Jetson Nano's dedicated GPU and hardware optimization for ML/AI tasks made it a more efficient platform for real-time anomaly detection. Despite these differences, both platforms handled DUDE-IDS operations effectively with minimal latency.

However, this study has certain limitations. First, while the LSTM-based DUDE-IDS demonstrated robust performance on test data, its scalability and effectiveness on larger, more diverse datasets have not been fully explored. Additionally, the computational demands of the system, particularly on resource-constrained devices like the Raspberry Pi, may limit its long-term feasibility for highly complex tasks. Finally, while the Jetson Nano showed better performance, its higher cost may be prohibitive for large-scale drone deployments. 

This comprehensive analysis highlights the potential of ML-based security frameworks to safeguard drones by effectively mitigating various attack vectors. The proposed DUDE-IDS framework stands as a promising solution for enhancing the cybersecurity and operational reliability of drones.



\bibliographystyle{ACM-Reference-Format}
\bibliography{bibliography.bib} 

\end{document}